\documentclass[twocolumn,a4paper]{article}

\usepackage{graphicx}
\usepackage{amsmath}
\usepackage{natbib}
\usepackage[caption=false,font=footnotesize]{subfig}

\advance \oddsidemargin by -\textwidth
\divide  \oddsidemargin by 2
\advance \oddsidemargin by -1in
\evensidemargin = \oddsidemargin

\newlength{\figurewidth}
\figurewidth = \textwidth
\advance \figurewidth by -\columnsep
\divide  \figurewidth by 2
\begin{document}

\title{An Introductory Review of Information Theory in the Context of
  Computational Neuroscience}

\author{Mark D.~ McDonnell\\
  Institute for Telecommunications Research\\
  University of South Australia\\
  Mawson Lakes, SA 5095, Australia\\
  \tt{Mark.McDonnell@unisa.edu.au}
  \and
  Shiro Ikeda\\
  The Institute of Statistical Mathematics\\
  Tokyo 190-8562 Japan\\
  \tt{shiro@ism.ac.jp}
  \and
  Jonathan H.~ Manton\\
  The University of Melbourne\\
  Victoria 3010 Australia\\
  \tt{jmanton@unimelb.edu.au}
}

\date{\today}

\maketitle

\begin{abstract}
  This paper introduces several fundamental concepts in information
  theory from the perspective of their origins in
  engineering. Understanding such concepts is important in
  neuroscience for two reasons. Simply applying formulae from
  information theory without understanding the assumptions behind
  their definitions can lead to erroneous results and
  conclusions. Furthermore, this century will see a convergence of
  information theory and neuroscience; information theory will expand
  its foundations to incorporate more comprehensively biological
  processes thereby helping reveal how neuronal networks achieve their
  remarkable information processing abilities.
\end{abstract}

\section{Introduction}
\label{sec:Introduction}

Norbert Wiener, the founder of cybernetics, wrote that it is the
``boundary regions of science which offer the richest opportunities to
the qualified investigator. They are at the same time the most
refractory to the accepted techniques of mass attack and the division
of labor''~\citep[p. 2]{bk:Wiener:cybernetics}. He went on to explain
that ``a proper exploration of these blank spaces on the map of
science could only be made by a team of scientists, each a specialist
in his own field but each possessing a thoroughly sound and trained
acquaintance with the fields of his neighbors; all in the habit of
working together, of knowing one another's intellectual customs, and
of recognizing the significance of a colleague's new suggestion before
it has taken on a full formal expression. The mathematician need not
have the skill to conduct a physiological experiment, but he must have
the skill to understand one, to criticize one, and to suggest one. The
physiologist need not be able to prove a certain mathematical theorem,
but he must be able to grasp its physiological significance and to
tell the mathematician for what he should look.''

Indeed, three giants of science, Wiener, von Neumann and Shannon,
realised in the 1940s the need for understanding the brain in terms of
the fundamental engineering principles applicable to any computational
device: energy, entropy and
feedback~\citep{bk:Wiener:cybernetics,bk:Neumann:brain,bk:Shannon:communication}.
This led to the Macy conferences (1946--1953) which attracted leading
scientists from across engineering and the physical and life sciences.
The Macy conferences were one of the earliest organised approaches to
transdisciplinarity and hailed by some as the most important event in
the history of science after World War II. They demonstrated the need
for, and the initial difficulties in, establishing a common language
powerful enough to communicate the intricacies of the relevant fields
across the physical and life sciences and engineering.

While their dream was not realised, this was primarily due to
insufficient experimental data. As time marched on, the barrier to
bringing together the ever more specialised disciplines grew
larger. With tremendous experimental advances having been made in the
past 60 years, it is timely to stand on the shoulders of these giants
and resume their quest. With this as motivation, the present article
endeavours to whet the appetites of neuroscientists and information
theorists alike for learning more of each other's fields.

\subsection{Relevance of Information Theory (and Feedback)}
\label{subsec:Relevance of Information Theory}

The human brain is often described as the most complex structure in
the known universe~\citep{j:Fischbach:brain}. Certainly, it is the
most efficient signal processing device known. Drawing only 20 watts
of power, the brain significantly outperforms engineered devices at
signal processing tasks such as source separation, feature extraction,
and speech and image recognition~\citep{Sarpeshkar}. This is all the
more remarkable because signals within the brain propagate very slowly
compared with those in a computer.

This suggests that the brain uses a paradigm for signal processing
very different from any developed in engineering. Why then should
engineering in general and information theory in particular have
relevance to understanding the brain? The answer lies partially in the
fact that engineers study fundamental laws pertinent to any system,
including biological ones~\citep{Berger.03,Sarpeshkar}. Indeed, John
von Neumann viewed the brain as a hybrid computer which performs
control, communication and computation, and concluded that information
theory is therefore essential for understanding its
functionality~\citep{bk:Neumann:brain}.  Wiener too recognised that
information theory was essential for a deeper understanding of
feedback and thus life~\citep{bk:Wiener:cybernetics}.  Anecdotal
evidence suggests Shannon himself, the father of information theory,
may have been partially motivated by how his brain processed
``information'' when performing a complex task such as juggling balls.

A few words on the concept of feedback are in order. Feedback refers
to achieving a task, such as keeping a car travelling at a constant
speed, by repeatedly measuring the current state, such as the car's
speed, and feeding those measurements back and using them to make the
requisite changes at the input, such as applying more or less pressure
to the accelerator of the car. Feedback is a fundamental concept in
engineering because it can militate errors caused by imprecisions and
external interference.

The brain too must use feedback to overcome
imprecisions~\citep{j:Burdet:nervoussystem,bk:Wiener:cybernetics,Marko,Todorov,Burdet,Franklin};
without feedback, we would fall over whenever we attempted to walk.
Within the sensory pathways there are tremendous numbers of feedback
paths connecting regions of higher-level brain function to regions of
lower-level functionality, giving rise to top-down processing theories
of the visual pathways and providing a mechanism for selective
attention.

Although they started out in different disciplines --- information
theory emerged from communication theory while feedback was studied in
control theory --- recent years have seen some convergence of feedback
and information theory. A fundamental question is what is the slowest
rate at which information must be fed back for the system to
work. Scientists have started to consider how fast the brain must be
processing information if we are able to walk properly and can move
our hand in a straight line even though random external forces are
impeding its motion in experiments~\citep{Burdet}. This is an example
of such convergence of two important theories.

By virtue of being introductory, the present paper focuses on
discussing information theory in ``one-way'' (i.e. feed-forward
exclusively) biological contexts. A comprehensive account of the brain
in engineering terms would necessarily involve the marriage of
information theory and control theory. Repeating the words of von
Neumann, the brain must be understood in terms of control (feedback),
communication (information theory) and computation.

\subsection{Information Theory and Communication}

It must be recognised that the mathematical discipline of
``information theory'' does not (and should not) capture all aspects
of how the word ``information'' is used in spoken language. Failing to
distinguish the two can lead to errors caused by flawed intuition in
one direction, or the inappropriate application of information theory
in the other\footnote{Blindly applying the mathematical equations
  defining entropy or channel capacity does not necessarily endow the
  resulting quantities with any meaning or validity; information
  theoretic quantities can be understood only with respect to the
  assumptions and limitations of information theory.}.

Information theory was invented in response to practical problems
faced by the designers of communication systems such as telephones and
data modems. The basic problem is to find an efficient way of
transmitting information from one place to another, whether it be a
probe sending information from the moon back to earth or a mobile
telephone sending and receiving voice and internet packets. Indeed,
consider the problem of one person trying to send a series of messages
to another person on the other side of a brick wall; for simplicity,
assume this second person is not allowed to speak or send any other
form of message to the first person such as an acknowledgement or
request for clarification (or ``retransmission''). How should the
first person send each message?

One thing is clear; the louder the first person shouts, the greater
the chance that the second person can understand the message over the
background noise (perhaps the neighbours are mowing their lawns).
Perhaps a little harder to appreciate but equally true in the digital
world, if the first person were to speak more slowly the second person
would have a greater chance of catching every word. The third
parameter that can be adjusted is the level of redundancy. When we
speak with a young child we tend to elaborate and use more words to
describe a concept in an attempt to increase the chances of correct
reception of the overall message.

In information theory, these parameters are referred to formally as
the transmission power, the transmission rate and coding (or
redundancy). The simplest form of coding is to repeat the message two
or more times. This is known as a repetition code.

Shannon's pioneering work shattered a long-held belief that with
finite power it was impossible to be able to transmit a message in
such a way as to \emph{guarantee} its \emph{perfect reception} even in
the presence of noise and other interference. Indeed, even if I
shouted at the top of my voice and repeated myself a hundred times,
every so often the interference (lawn mowers?) will prove too great
and my message will be lost.

The answer lies in coding; repetition codes are not particularly good
codes. Shannon realised that there exist very clever codes which can
ensure that any two messages are so different from each other that the
receiver can correctly decide which message was sent despite the
interference. Technically, perfect reception requires the receiver to
listen forever before deciding which message was sent but the key
point is that given any positive but arbitrarily small probability of
error (such as one message being incorrectly received in $10^{12}$
messages) then a code can be constructed which achieves this level of
performance in finite time, and more importantly, the transmission
power does not need to be increased. Increasing transmission power to
achieve a particular error rate is grossly inefficient compared with
choosing a better code. In the example of one person trying to convey
a message to the other person, the secret is to share a codebook
beforehand, and a different sequence of sounds, one for each message
that may be sent, is written in it. ``It will rain tomorrow'' might be
encoded as (a segment of) Beethoven's 5th symphony while ``It will be
sunny tomorrow'' might be encoded as a hard rock song. These two
encoded messages are ``sufficiently different from each other'' to
have very little chance of being confused. More importantly, any small
fragment of the two messages are different. This is how interference
is overcome. Because interference itself has limited power (otherwise
the game would not be fair!) then even if there are times when the
interference is particularly bad, there will be other times when the
interference is back to normal and in the long run, there is no
confusing Beethoven for hard rock.

For the transmission rate to be acceptable the codebooks would need to
contain more than just two messages. (With two messages, each one
encoded by a five-minute song, the transmission rate would be 1 bit
per 5 minutes.) In the same way that the transmission power does not
need to go to infinity, the transmission rate need not go to zero.
Precisely, Shannon discovered a quantity known as the channel
capacity. If the transmission rate is less than the channel capacity
then communication with any desired level of accuracy is possible
whereas if the transmission rate exceeds the channel capacity then it
becomes impossible to have arbitrarily good performance in finite
time. As is to be expected, the channel capacity depends on the
interference. The more destructive the interference the lower the
channel capacity.

\subsection{Intra-organism communication}

Messages are also passed around within an organism. Information
gathered by an organism's senses must be communicated for it to have
any effect on the organism. Within the brain and nervous system,
information is manipulated in at least three different ways:
\begin{itemize}
\item information acquisition (sensory transduction);
\item communication between spatially separated regions (information
  transmission);
\item memory formation and recall (information
  storage~\citep{Varshney.06}). 
\end{itemize}
In brains, each of these are essential for the emergence of broader
functions that might be termed `computation'. Communication is perhaps
most fundamental, since information from the senses needs to be
communicated in order for it to have any affect on an organism, while
information storage, whether in computers or brain memories, can be
viewed as communication from the past to the present or future.

Understanding how the brain stores and transmits information is
tantamount to understanding the brain as a whole because if we could
``listen in'' to brain messages it would surely be just a matter of
time before we understood the computational side, too. That said, the
possibility that the brain does not separate out information
transmission from information processing must be considered. Whereas
computer architectures have mechanisms known as buses for moving
information between different processing units, the brain might take a
more efficient distributed approach and simultaneously process and
communicate in a nonseparable fashion. It has been stated that
``computation in the brain always means that information is moved from
one place to another''~\cite[p. 116]{Buzsaki}. A comprehensive
understanding of the brain's mechanisms for internal communication
will likely form an integral part of more advanced theories about how
`computation' arises within brain networks.

Regardless of how the brain actually processes information, at the end
of the day the brain is an input-output system (we react to what we
sense) and therefore subject to the same laws as any other
input-output system. Information theory is therefore relevant to
understanding how the brain works, and conversely, it is highly likely
that advances in the field of information theory will be made in
synergy with new discoveries of the computing paradigms used by the
brain~\citep{Cohen.04,Sarpeshkar,Berger.03}. Indeed, information
theory may have to expand to address new neurobiologically relevant
questions if it is to be powerful enough to explain all aspects of how
the brain manipulates information.

To demonstrate the relevance of even simply thinking in information
theoretic terms, Landauer estimated that humans learn information at a
rate of about two bits per
second~\citep{j:Landauer:memory}\footnote{To put this in perspective,
  a digital camera typically stores a single photo using around
  10,000,000 bits. Clearly then, we extract only a very small amount
  of information from what our senses receive.}. Taking memory loss
into account, a person will accumulate approximately two billion bits
of information in a lifetime (or approximately 240MB in computing
terms). Since our brain has many more synapses than two billion,
Landauer concludes that ``possibly we should not be looking for models
and mechanisms that produce storage economies but rather ones in which
marvels are produced by profligate use of capacity.''

\subsection{Outline of Paper}

According to~\cite{Cohen.04}, biological science asks six kinds of
questions about domains ranging from molecules and cells, up to the
biosphere:
\begin{enumerate}
\item How is it built? (Structures)
\item How does it work? (Mechanisms)
\item What is it for? (Functions)
\item What goes wrong? (Pathologies)
\item How is it fixed? (Repairs)
\item How did it begin? (Origins)
\end{enumerate}
Utilising information theory in neuroscience is ultimately useful
only if it can address one or more of these questions.

In this paper we advocate that information theory 1) can be a useful
framework for finding answers to some of these questions; but 2) must
be broadened for its theorems to be directly applicable to neuronal
networks. Although information manipulation can happen at very
different levels of organisation, such as storage of information in
genes, or communication at the level of synaptic transmission between
cells, or at that of spiking patterns of neurons in a network, in this
paper we will be focusing on examples that involve spiked-based
communication between neurons.

In making these points, it is necessary for us to introduce the most
basic and well-known information theoretic concepts in
Section~\ref{sec:info}, before discussing the challenges of applying
the theory meaningfully to questions in neuroscience in
Section~\ref{S:neuro}. Then in Section~\ref{sec:NeuralCapacity} we
summarise a specific example which illustrates that information
theoretic approaches depend critically on different assumptions that
could be made about neural systems.  Finally in
Section~\ref{S:conclusion} we conclude the paper with some closing
remarks on the material we cover and briefly summarise recent
developments on information theoretic approaches in neuroscience that
extend well beyond the classical ideas we present, thereby with
increasing relevance to neurobiological systems.

\section{The basics and utility of Shannon Information
  Theory}\label{sec:info}

This section briefly explains key concepts from Shannon information
theory and hints at possible contributions in neuroscience. By {\em
  Shannon information theory} we are referring to a specific sub-part
of the broader field of {\em information theory}. The latter, by
definition, encompasses any {\em mathematical theorems about
  information}, and therefore is not confined to well-known concepts
introduced by Shannon, such as entropy and mutual information. As we
discuss later, information theory beyond Shannon theory may be very
important in neuroscience.

Shannon's milestone paper~\citep{Shannon.1948} that founded the field
of information theory showed to the world that introducing the right
kind of redundancy was the key to moving information from one place to
another in an efficient and reliable manner. Since information sources
such as spoken voice or PDF (portable document format) documents
generally contain the wrong kind of redundant information, Shannon
proposed a two-step process: first remove the existing redundancy by
compressing the message to be sent, then introduce the right kind of
redundancy for communicating the message through the channel at
hand. These two important concepts are known as ``source coding'' and
``channel coding'' respectively. They motivate several fundamental
questions including determining the maximum amount of compression
possible of an information source. Answers to these questions are
given in terms of quantities such as \textit{entropy} and
\textit{mutual information}. It is important to realise that these
quantities were given special names because they serve to answer
important questions for a particular class of problems. It would be a
mistake to assume without additional justification that they are
applicable or even meaningful beyond the bounds of the original
questions for which they serve as the answers to. See
e.g.~\cite{Johnson.08ITW} for more discussion.

\subsection{Entropy and Source Coding}
\label{subsubsec:entropy}

\begin{figure}[htb]
  \centering
  \includegraphics[width = 0.9\figurewidth]{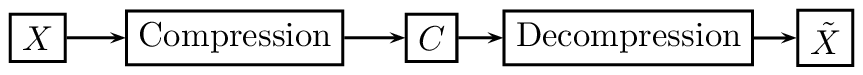}
  \caption{Source Coding.}
  \label{fig:sourcecoding}
\end{figure}

Living in the digital age, readers will be familiar with compressing
files. Zipping up a file to send to a friend is an example of lossless
compression. Generally (but not always) the compressed file will be
smaller than the original yet no information has been lost; the friend
can recover the original file by decompressing the compressed file
(Fig.~\ref{fig:sourcecoding}). For compressing music or photos,
significantly greater compression can be achieved by using lossy
compression algorithms such as MP3 and JPEG. As the name suggests,
some information is lost~\citep{Berger.98}. The original can be
recovered sufficiently well for a satisfactory compromise to have been
reached; a small amount of quality is sacrificed for a large saving in
storage space.

The remainder of this section discusses lossless compression only.
Consider the problem of compressing a short message of length 8 bits.
A bit is simply a ``{\tt 0}'' or a ``{\tt 1}'' so an 8-bit message is
a sequence of eight zeros and ones, such as ``{\tt 01011101}'' or
``{\tt 10101010}''\footnote{Of course, any other alphabet could have
  been used. An 8-character message consisting of a string of 8
  letters of the alphabet might look like ``{\tt abzpuikq}'' and the
  same reasoning would apply.}. A calculation (or by writing out all
the possibilities if need be, starting with ``{\tt 00000000}'', ``{\tt
  00000001}'' and continuing until ``{\tt 11111111}'') shows that
there are precisely 256 different 8-bit messages. Compressing a
message would mean using fewer than 8 bits to store the message.  A
simple enumeration shows that this is impossible as stated; there are
only 128 7-bit messages, not enough to represent all possible 8-bit
messages. How then does a computer compress a file losslessly?

The secret is that there is often redundancy in the kinds of
information that people are interested in. Equivalently, it is
generally the case that not all messages have an equal chance of
occurring. For argument's sake, assume that out of the 256 possible
messages, there are 15 messages which occur most of the time. To
exploit this, we may decide to use 4 bits to represent each of these
messages. Precisely, ``{\tt 0000}'' would represent the first message,
``{\tt 0001}'' the second, up to ``{\tt 1110}'' for the 15th
message. To represent any other message, we would first write down
``{\tt 1111}'' to mean ``not one of the 15'' and then we would write
down the original message using 8 bits.  This means that 15 of the
messages can be written down using only 4 bits but the remaining
$256-15=241$ messages now require $4+8=12$ bits for their storage.

The only way to make this meaningful is to consider repeating this
compression exercise many times. If we had to store a very large
number $N$ of 8-bit messages using this scheme, how many bits will be
required?  Assume that $K$ out of the $N$ messages belong to the set
of 15 special messages. These $K$ messages require 4 bits while the
remaining $N-K$ messages require 12 bits, or in total, $4K+12(N-K)$
bits are required compared with $8N$ bits had we not compressed the
messages.  Provided $K$ is sufficiently large, we will have succeeded
in compressing the data. For example, if $K=750$ and $N=1000$ then we
would require only 6,000 bits rather than the original 8,000 bits.

The conventional way to describe the above scenario is to work with
probabilities. We assume that the messages we are being asked to
compress are being generated at random and there is no correlation
between the message we are being asked to compress now and the
messages we have already compressed. Mathematically, we represent the
original sequence of messages by a sequence of independent and
identically distributed random variables $\{x_1,x_2,\cdots\}$, each
having a probability density $p(X)$. In the above example, each $x_k$
would be an 8-bit message (or equivalently, a number between 0 and 255
inclusive) and $p(X)$ would be the probability that a particular
message $X$ is chosen. For concreteness, assume that each of the first
fifteen messages have a 5\% chance of occurrence (meaning there is a
75\% chance of a randomly chosen message being one of these 15 and
thereby corresponding with the earlier choice of $K=750$ and $N=1000$
). Then $p({\tt 00000000})=p({\tt 00000001})=\cdots=p({\tt
  00001110})=0.05$. We assume that all other messages each have a
probability of ${0.25}/(256-15)$ of occurring. The expected number of
bits required to compress a single message can then be calculated by
summing over $i$ the probability that the $i$ th message occurs
multiplied by the number of bits required to represent the $i$ th
message. When most of the probabilities are the same the calculation
simplifies. With the values given above, the expected length is
calculated to be
$15\times0.05\times4+(256-15)\times\frac{0.25}{256-15}\times12=6$.
Thus, on average, $6N$ bits would be required to compress $N$ messages
drawn at random if the above scheme were used.

Is there a better compression scheme, one which requires fewer than
6 bits per message on average? In fact, what is the best possible?
As elucidated presently, Shannon was able to answer these questions.
First though, a technicality needs mentioning.

Coding each message separately, as was done above, is inefficient.  It
is better to concatenate a series of messages and compress them all at
once; this provides more opportunity for better compression through
the simple fact that there are more compression schemes to choose
from. (It also alleviates the wasted space caused by otherwise having
to use an integer number of bits to represent each message.)

It is therefore quite standard to refer to each $x_k$ as a symbol
rather than a message and ask how many bits per symbol on average must
be used to compress the infinitely long sequence of independent and
identically distributed symbols $\{ x_1,x_2,\cdots\} $ if each symbol
has a probability $p(X)$ of occurring\footnote{It is traditional in
  probability theory to use an upper case letter to represent the
  random variable while the corresponding lower case letter represents
  realised values.}.

When $X$ is a random variable and its distribution is $p(X)$, its
\emph{entropy} is defined as
\begin{align}
  H(X)=-E_{p(X)}[\log p(x)],
  \label{eq:entropy_discrete}
\end{align}
where $E_{p(X)}[\cdot]$ denotes the expectation with respect to
$p(X)$. The practical operation is a summation when $X$ is discrete
and an integration when $X$ is continuous. When $2$ is the base of
logarithm, i.e. $\log_2$, the units of entropy are bits and $H(X)$ is
precisely the number of bits per symbol required on average to
compress an infinitely long sequence of symbols when each symbol has
probability $p(X)$ of occurring.

It is for this reason that people endeavour to explain entropy as
quantifying the ``ambiguity'' or ``uncertainty'' about the random
variable $X$. When $X$ has only one possible state (that must
therefore occur with probability $1$), there is no ambiguity about $X$
and the entropy is $0$. However if $X$ takes one of two states with
probability $p$ and $1-p$, respectively, ($0\le p\le1$), entropy is
maximised when $p=0.5$ and $H(X)=\log_22$. This is exactly 1 bit and
implies that a sequence of equally likely zeros and ones cannot be
compressed. Note also that if $p=0$, $0\log_20=0$.

\emph{Shannon's source coding theorem} states that (in the limit as
the number of symbols goes to infinity) it is \emph{possible} to
compress each symbol to $H(X)$ bits on average (and \emph{impossible}
to do better). It does not however say how to design such a source
code.  Furthermore, the practical construction of compression and
decompression methods is complicated by considerations of algorithmic
efficiency (which affects battery life in portable equipment such as
mobile telephones) and latency (how long the receiver must wait from
the time a symbol is sent until that symbol can be received and
decoded). That said, having a target to aim for is extremely useful
and entropy provides that target for source compression.

The reader may wish to verify that for the example introduced in this
section, the corresponding entropy is $5.72$ bits per symbol. This
represents the best any compression scheme can hope to achieve, and
indeed, it is lower than the 6 bits per symbol scheme presented here.

\subsection{Mutual Information, Channel Capacity and Channel Coding}

The following example of a binary symmetric channel will be used to
add concreteness to the ensuing introduction of \emph{mutual
  information} and \emph{channel capacity}. Let $\{s_1,s_2,\cdots\}$
denote a binary sequence which is to be transmitted to another person
or device. It is called the source sequence. The medium through which
a message can be sent from one person or device to another is called
the channel. Mathematically, a channel takes a sequence at its input
and it generates another sequence at its output. If the channel were
ideal, it would simply copy its input to its output and communication
would be straightforward. Generally though, the channel is not ideal.
It introduces random errors. If $\{x_i\}$ is the binary input sequence
(which is shorthand notation for $\{x_1,x_2,\cdots\}$) then the binary
output sequence $\{\tilde{x}_{i}\}$ of a binary symmetric channel with
error probability $p$ is given by the rules that 1) for each integer
$i$, the output $\tilde{x_{i}}$ at time $i$ depends only on the
corresponding input $x_i$ at the same time $i$; and 2) the probability
that $\tilde{x}_i$ differs from $x_i$ is $p$.  If $p=0.1$ then on
average one in every ten symbols will be corrupted, meaning either a
$0$ was sent and a $1$ was received, or a $1$ was sent and a $0$ was
received.

\begin{figure}[htb]
  \centering
  \includegraphics[width=0.6\figurewidth]{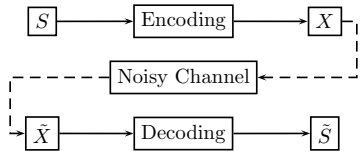}
  \caption{Channel Coding.}
  \label{fig:channelcoding}
\end{figure}

What sequence $\{x_i\}$ should be sent over the channel if the
ultimate aim is to send $\{s_i\}$ reliably to the receiver, assuming
of course that the receiver can process the output $\{\tilde{x}_i\}$
of the channel before deciding what it believes the message $\{s_i\}$
is? This is illustrated in Fig.~\ref{fig:channelcoding} where the
operation of generating $\{x_i\}$ from $\{s_i\}$ is called (channel)
encoding and the operation of generating $\{\tilde{s}_i\}$, the
receiver's best guess at the original message, is called (channel)
decoding.

For simplicity, often the encoding and decoding processes work on
blocks of data. Precisely, the original source sequence $S$ is divided
up into subsequences of length $K$. Each of these is encoded to a
longer binary sequence $X$ with length $N$. For example, a simple
$K=2$, $N=3$ block code would be to add a parity bit (i.e. a bit that
is zero when the sequence has an even number of zeros, and a one when
an odd number) after every two symbols, so: ``$\tt{00}$'' becomes
``$\tt{000}$''; ``$\tt{01}$'' becomes ``$\tt{011}$''; ``$\tt{10}$''
becomes ``$\tt{101}$'' and ``$\tt{11}$'' becomes ``$\tt{110}$.''
Therefore, the sequence ``$\tt{0111}$'' becomes ``$\tt{011110}$''
where the 3rd and 6th bits are the introduced parity bits.

This coded sequence is transmitted through the channel. At the other
end, the receiver reverses the process, converting each block of $N$
symbols back into a block of $K$ symbols. In this particular case,
introducing just a single parity bit does not allow the receiver to
have a better guess at what the original message is, but it does allow
the receiver to detect if a single bit has been changed. This is
called error detection. Error correction, when the receiver is not
only able to detect an error has occurred but can fix the error and
therefore recover the original message, requires more redundancy to be
introduced, that is, choosing $N$ to be larger than $K+1$. (If there
are too many errors then error correction would fail, but the key
point is that the probability that several consecutive bits are wrong
is significantly smaller than if a single bit were wrong, therefore a
small increase in redundancy allows a substantial increase in
reliability.)

The two lengths $K$ and $N$ together define the ``rate'' of the code,
which perhaps is better understood as measuring the decrease in
throughput caused by the introduction of redundancy by the encoder.
Precisely, in the above example, the rate of the code is $R=K/N$,
meaning that if the channel can accept encoded symbols at a rate of 1
bit per second then the source symbols must have a rate of only $R$
bits per second.

Reducing the rate enables more redundancy to be introduced which can
be used to increase the chance of the receiver being able to work out
what message was sent. Shannon's remarkable observation was that there
is a much better way of increasing the chance of correct reception
than by decreasing the rate towards zero. For a fixed rate $R$, the
block size $K$ can be increased (thereby increasing $N$ according to
the formula $N=KR$~). This allows a more sophisticated form of
redundancy to be introduced (but at the price of introducing greater
latency; the receiver must receive $N$ symbols before it can work out
what the corresponding $K$ message symbols were).

Shannon proved that there exists a rate $C$, called the channel
capacity, such that for any rate $R$ strictly less than $C$ and any
desired error rate $\epsilon>0$ (meaning that the probability that the
receiver decodes a bit incorrectly is less than $\epsilon$, which
might be chosen to be $\epsilon=10^{-9}$ or smaller in practice),
there exists a $K$ (possibly quite large) and a block encoder and
decoder pair such that the receiver can correctly decode each bit of
the source message with error probability less than $\epsilon$. This
is customarily summarised by saying that error-free communication is
possible at rates below the channel capacity\footnote{Shannon also
  proved the converse, that no scheme (even non-block-based coding
  schemes) can achieve arbitrarily small error rates if their rate is
  greater than or equal to $C$.}.

Shannon was able to give a formula for computing the channel capacity
$C$. When this formula (described below) is applied to the above
example of a binary symmetric channel with probability of error $p$,
the channel capacity is found to be $C=1+p\log_2p+(1-p)\log_2(1-p)$,
meaning for example that if the channel can transfer one bit per
second then the source symbols must arrive slower than $C$ bits per
second.  If $p=0.1$ then $C=0.531$ meaning that for every 1,000 source
symbols, just over 1,883 encoded symbols are required for reliable
communications.

The formula for channel capacity involves a quantity called
\emph{mutual information}. Intuitively, the mutual information of the
input and the output of the channel measures how much information the
output provides about the input; the more reliable the channel the
higher the mutual information. It is therefore reasonable to expect
that the larger the mutual information the greater the channel
capacity.

Bearing in mind that ``information'' is a very general word and it is
therefore not possible to capture all its nuances in a single
mathematical definition, it is expedient to return to the idea in the
previous section of using asymptotic compressibility as a measure of
information. It turns out that this is the right definition to use
when it comes to determining the capacity of a channel (which in
itself is an asymptotic measure).

Suppose there are two random variables, $X$ and $Y$, and they are
somehow related to each other. For example, $X$ might denote
temperature while $Y$ denotes humidity. Even simpler, $X$ might
represent the outcome of rolling a 6-sided die while $Y$ is given the
value ${\tt 0}$ if the die landed on an even number, or ${\tt 1}$ if
odd. Knowing $Y$ gives partial information about $X$; how can we
measure how much information $Y$ tells us about $X$?

The fact that $Y$ gives partial information about $X$ is reflected in
the fact that if $Y$ is known then $X$ can be compressed more than if
$Y$ were not known. In the above example, if $Y$ were not known then
it is impossible to compress $X$ because each of the outcomes is
equally likely; we are forced to use one of six possible symbols (or
$\log_{2}6$ bits) to store each sample of $X$; the entropy of $X$ is
$H(X)=\log_{2}6$. If $Y$ is known though then only one of three
possible symbols (or $\log_2 3$ bits) needs to be stored; the
\emph{average conditional entropy} is $H(X|Y)=\log_2 3$. The
additional amount of compression possible, $I(X;Y)=H(X)-H(X|Y)$, is
called the mutual information and measures the amount of information
$Y$ provides about $X$. It turns out that mutual information is
symmetric --- $I(X;Y)=I(Y;X)$ --- hence there is no need to specify
the order of $X$ and $Y$. In the above example, if $X$ is known then
$Y$ is known, therefore no additional bits are required to store $Y$
if $X$ is known: $H(Y|X)=0$. Since $H(Y)=\log_2 3$ it is indeed the
case that $I(Y;X)=H(Y)-H(Y|X)=\log_2 3=I(X;Y)$\footnote{The reason for
  this symmetry is that if we were to compress $X$ first then compress
  $Y$, or if we were to compress $Y$ first then compress $X$, we end
  up either way with having compressed optimally the joint sequence
  generated by $X$ and $Y$. Mathematically,
  $H(X,Y)=H(X)+H(Y|X)=H(Y)+H(X|Y)$ from which it follows immediately
  that $I(X;Y)=I(Y;X)$.}.

Returning to the channel capacity calculation, assume that a sequence
generated by $X$ is sent through the channel. The output sequence is
itself generated by a random variable, call it $Y$. If the receiver
wants to recover $X,$ it needs at least an extra $H(X|Y)$ bits of
information (for otherwise there would be an even more efficient
scheme for compressing $X$ than the best possible, a
contradiction). Looking at it from another angle though, this implies
that $I(X;Y)=H(X)-H(X|Y)$ bits of information have somehow been
transmitted successfully with each use of the channel (since with an
extra $H(X|Y)$ bits of carefully chosen information it is
theoretically possible to recover $X$).  For the case of the binary
symmetric channel with error probability $p$, a reasonably
straightforward calculation shows that if the input $X$ takes the
value $1$ with probability $q$ and the value $0$ with probability
$1-q$ then the mutual information of the input $X$ and the output $Y$
is $I(X;Y)=H(q)-H(p)$ where
$H(\theta)=-\theta\log_{2}\theta-(1-\theta)\log_{2}(1-\theta)$ is the
number of bits required to compress a binary sequence taking the value
$1$ with probability $\theta$ and the value $0$ with probability
$1-\theta$.

Although we must have the channel input $X$ represent the source
message $S$ in some way, there is otherwise arbitrary freedom in how
to choose $X$. Why not choose $X$ to maximise the mutual information?
The largest value $H(q)$ can take is $1$ (which occurs when $q=1/2$~).
Therefore, the largest number of bits we can ever expect to transmit
reliably through the binary symmetric channel is $1-H(p)$ bits per
usage of channel. If $p=0.1$ then $1-H(p)=0.469.$ In this case, at
most every $0.469$ bits of the source message must be expanded to $1$
bit (since the channel transmits 1 bit per usage), or in other words,
we must have the rate of the code (see above) satisfy $K/N<0.469$.
Remarkably, Shannon proved that this bound is achievable; whenever the
rate is less than the maximum of the mutual information, (as close as
you like to) error-free communication is
possible\footnote{Note that Shannon proved ``it is
    achievable'' in the limit when $N$ and $K$ goes to infinity, but
    did not show ``how to achieve it.'' In order to be close to the
    bound, we generally need a good error correction code and $N$ and
    $K$ must be very large.}.

It is of interest to note that while here we have considered an
example where $X$ is a discrete random variable, the most well known
case of a channel for which the capacity achieving input distribution
is known, is the additive Gaussian noise channel, with a power
constraint on the input. In this case, the capacity achieving input
distribution is in fact continuous, i.e. a Gaussian distribution. As
we discuss later though, it is far more common for the capacity
achieving input to be discrete.

We now summarise and precisely define the important information
theoretic terms that we have introduced and discussed above without
stating their formal definitions. Each of these are defined
mathematically as follows. We already introduced entropy, in
Eqn.~\eqref{eq:MI_discrete}. The average conditional entropy requires
a double expectation:
\begin{equation}
  H(Y|X)=-E_{p(X)}\bigl[E_{p(Y|x)}[\log p(y|x)]\bigr].
  \label{eq:condentropy_discrete}
\end{equation}
As an aid to intuition, consider a single outcome of the random
variable $X.$ The entropy of $Y$ given $X$ can be calculated from
Eqn.~\eqref{eq:condentropy_discrete} by calculating the expectation
with respect to the conditional distribution of $Y$ given $X=x$ . If
this is carried out for all possible outcomes of $X$, the result is a
function of $x$. This function can then be averaged with respect to
the distribution of $x$, and by definition, the result is the average
conditional entropy, $H(Y|X)$.

As mentioned above, the mutual information can be expressed as $I(X;Y)
= H(X)-H(X|Y) = H(Y)-H(Y|X)$ . In what follows below we write mutual
information in a different form based on another entity called
\emph{relative entropy} or \emph{Kullback-Leibler divergence}. This is
defined as
\begin{equation*}
  D(p(X)||q(X))=E_{p(X)}\left[\log\frac{p(x)}{q(x)}\right],
\end{equation*}
where $p(X)$ and $q(X)$ are two distributions of the same random
variable $X$. Note that the relative entropy is positive and is equal
to 0 if $p(X)$ is identical to $q(X)$. Mutual information is defined
as the relative entropy between the joint distribution of $X$ and $Y$,
and the product of the marginal distributions of $X$ and $Y$:
\begin{align}
  \begin{split}
  I(X;Y)&=D(P(X,Y)||P(X)P(Y))\\
  &=E_{p(X,Y)}\left[\log\frac{p(x,y)}{p(x)p(y)}\right].
  \end{split}
  \label{eq:MI_discrete}
\end{align}
It is straightforward using $p(x,y)=p(y|x)p(x)=p(x|y)p(y)$ to obtain
the above stated relationships between mutual information and
entropy. The definitions as written here hold for both discrete and
continuous distributions of $X$ and $Y$. In this section we have
considered only a simple discrete case, where $X$ and $Y$ are both
binary. In general they can have any number of states, or be
continuous, as is the case below. In full generality, the channel
capacity is defined as
\begin{align*}
  C & =\sup_{P(X)}I(X;Y).
\end{align*}

\section{Challenges of Utilising Information Theory in Neuroscience}\label{S:neuro}

In this section, some of the challenges of integrating information
theory into neuroscience are touched upon. In particular, we must make
assumptions about the way in which information is represented in the
brain, whereas in engineering this is specified by the
designer. Ultimately it will be necessary to extend the frontiers of
information theory if it is to encompass in its entirety the
information processing techniques of neuronal networks. Such an
expansion would involve in part the greater integration into
information theory of systems and control theory from engineering and
the theory of computation from mathematics. Whereas engineers aim to
keep separate communication circuitry from computation circuitry so as
to simplify the design and analysis of engineered systems, there is no
reason why nature should maintain such a separation. Evolution tends
to find efficient designs and not necessarily ``simple'' designs.

It would be counter-productive though to assume that information
theory in its current form could not be applied usefully in
computational neuroscience. One place it is immediately applicable is
the early sensory pathways where information is primarily flowing in
one direction. Considerably extra care must be taken when feedback
loops are present. This is especially the case because (in
experiments) we have control over the input signal itself and hence
can investigate how a known signal is communicated from one neuron to
another. The complication though is that it appears the information
is being processed at the same time it is being communicated.

The brain heavily compresses the information it receives from its
sensory systems. Since the entropy of a signal determines precisely
how much (lossless) compression is possible, it sets fundamental
limits which must be respected by any system, including biological
systems. It is no surprise then that the estimation of entropy of
neural signals based on experimental data is an active research
area~\citep{Borst.99,Panzeri.07,Vu.09}.

In the brain, neurons communicate with each other and transfer
information.  The primary means of communication are the spikes of
each neuron~\citep{Rieke}, and it is parsimonious to model their
occurrences as depending randomly on the neuron's
input~\citep{Poggio.64,Mainen.95}. Thus, neurons communicate through a
noisy channel, and mutual information should therefore play an
important role in understanding the nervous system and brain

In order to consider a neuron as a communication channel, we need to
consider what we mean by ``communication'' in the specific context of
biological neurons. There are several important concepts to consider
before we can begin to discuss a specific example of the application
of information theory in neuroscience.

\subsection{Communication Channels and Modulation}

A definition of communication requires the existence of a physical
medium that allows propagation of energy from one place (an ``energy
source'') to another place where that energy has some causal effect
(an ``energy sink'')%
\footnote{Communication can also take place from the past to the
  future, in a fixed location, such as when writing to a memory device
  then reading it back again at a later date, but here we focus on
  place-to-place communication. %
}. We also need to define a means by which some property of the source
can be altered in a way that results in an observable difference at
the sink after propagation through the channel. In communications
engineering theory, the energy propagation is called ``transmission,''
the source is known as a ``transmitter'' and the sink as a
``receiver.''  These concepts are not sufficient for
communication. There also needs to be an ``information source'' that
is initially observable at the transmitter's location, but not at the
receiver's. Communication requires the transmitter to alter the energy
source in a manner that reflects the information source, and that can
subsequently be observed at the receiver after propagation. This
conversion from information source to energy source is known as
``modulation.''

A familiar example where each of these concepts is readily identified
is analog AM or FM radio transmission, in which recorded sound signals
are communicated, and then reproduced via a speaker. In this example,
the transmission medium can be a vacuum\footnote{Vacuum is thought of
  a transmission medium for electromagnetic waves.} or air, the
propagating energy source is electromagnetic radiation, and the
transmitter modulates the electromagnetic waves in a manner that
reflects the recorded sound signal. AM is amplitude modulation, and
means that a single frequency sinusoidal wave of E-M (electromagnetic)
radiation has its amplitude changed over time. FM is frequency
modulation, which means the amplitude remains constant, while the
carrier frequency is changed over time.

\subsection{Neuronal Spikes and Spike Interval Coding}

Modulation of the energy source can be thought of as a code, since it
requires a conversion from one kind of information representation to
another. Indeed, in neuroscience, modulation has a more general
meaning than in communications engineering, and the conversion from an
information source to variations in a parameter of the energy source
is instead known as a ``code.'' This is largely in contrast with
communications engineering, where ``code'' instead refers to
conversion between different representations of the information source
prior to transmission at the source, for example ``source coding'' and
``error correction coding.''

If we wish to consider communication between neurons, we need to
identify the transmission medium, the form of energy propagation, and
a modulation mechanism. From now on we will use neuroscience
terminology, and refer to modulation as the ``code.'' Further, we will
refer to the information source as the ``input,'' and the observable
effect at the receiver that results from the input as the ``output.''

Although over longer time scales the plasticity of neurons can
encode/carry information, in shorter time scales the primary physical
medium for communication seems to be the axons of neurons, and the
energy propagation is a pulse-like wave of voltage that travels along
an axon where it may be received by other neurons at synaptic
junctions. These pulses are known as action potentials, or
spikes. Typical cortical neurons transmit spikes to many other
neurons, and receive spikes from many neurons.

While there are a number of different ``communication channels'' in
neuronal circuitry --- including segments of the dendritic tree which
carry post-synaptic potentials towards the soma of the cell --- we
choose to focus on action potentials because it is
one of the most important communication mechanisms between two neurons.

The other concept we must also attempt to identify is the way in which
spikes are coded (modulated) in order to communicate information.  Two
possibilities are the height and the width of each spike. However,
these are observed to be close to identical in most cases, and do not
seem to be information carrying parameters. Instead, it is the
interval between spikes (ISI: inter-spike interval) that is thought to
play an important role in carrying information through a neuronal
channel.

\begin{figure}[htb]
  \centering
  \subfloat[Temporal coding.]{
    \includegraphics[width = 0.47\figurewidth, clip]{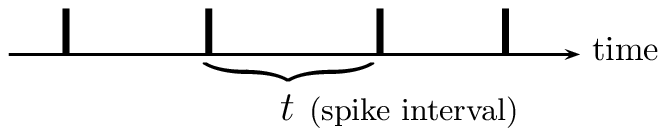}
    \label{fig:TC}
  }
  \subfloat[Rate coding.]{
    \includegraphics[width = 0.47\figurewidth, clip]{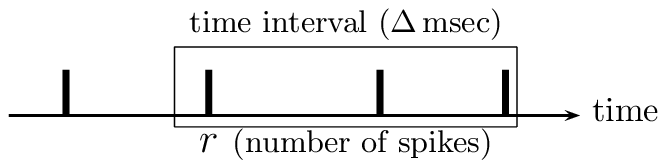}
    \label{fig:RC}
  }
  \caption{Two forms of spike codes.}
  \label{fig:TCandRC}
\end{figure}

Given this, how do ISIs represent information? In neuroscience there
are mainly two different ideas. One idea is that the ISI itself (see
for example~\cite{MacKay.52}) carries information. This is called
``temporal coding'' (Fig.~\ref{fig:TC}). The other is that the number
of spikes in a fixed time interval (see~\cite{Stein.67,Lansky.04})
carries information. This is called ``rate coding''
(Fig.~\ref{fig:RC}).

So far we have only stated that an input can be communicated to a
receiver output. If this is a perfectly repeatable process, the rate
at which information can be transmitted depends on the rate at which
the input is updated, and --- in line with Section~\ref{sec:info} ---
also depends on the probability distribution of the input, via its
entropy. The transmission is usually not perfect, and noise is
introduced. This fact leads us to consider the information theoretic
concepts of mutual information and channel capacity.

Information theory is not concerned with the type of modulation. It
requires an abstraction that specifies only what the observable output
variable should be. Since we are not designing a system, we must make
some guesses about aspects of the input and output for a neuronal
communication channel, and then proceed to calculations of mutual
information.

Therefore, in Section~\ref{sec:NeuralCapacity} where we consider the
channel capacity of a neuron model, we necessarily begin by
specifically defining the input and output of the channel, and state a
model for the channel noise.

\section{Example: Channel Capacity of a Neuron}
\label{sec:NeuralCapacity}

In this section we present some of our results on the channel capacity
for simple neuron models.  One reason for providing this example, is
to illustrate that there is no simple single formula for channel
capacity, and hence assumptions about the underlying model are very
important. If these assumptions change, the channel capacity also
changes.

As we have seen, channel capacity is the maximum amount of information
that can be transferred through a noisy channel in a unit time. It may
be much larger than the actual information transmission rate. This
brings us to a natural question, that is, why do we need to know the
capacity?

Channel capacity is something similar to the maximum speed indicated
in the speedometer of an automobile. While you will likely never drive
with that speed, the maximum speed is useful because it tells you the
potential of the automobile, even though you drive with moderate
speed. Channel capacity provides not only the upper limit of the
possible information transmission rate, but also describes how good
the channel is.

Although there is much interest in the quantity in
neurophysiology~\citep{Borst.99}, theoretical work is
rare~\citep{MacKay.52,Stein.67,Johnson.10,Suksompong.10}. We have
obtained some interesting results on the capacity from two different
viewpoints. The details will be given below.

\subsection{Inputs and Noise of Channel}
\label{subsec:Inputs and Noise of Channel}

We consider here a single spiking neuron, and assume that the input to
the neuron controls the expectation of the neuron's output ISIs.
Using the terminology introduced above, the information source
modulates the ISI. We introduce channel noise to the picture by
assuming that the ISI is a gamma-distributed random variable, when the
input to the neuron remains constant.

Biologically, each cortical neuron receives inputs from a lot of
(pre-synaptic) neurons and each sensory neuron receives physical
stimuli. The above assumption is to model all the inputs to the neuron
as a single parameter $\theta$. Although this assumption may seem too
simple, $\theta$ is a time varying function and is able to represent a
a lot of possible functions. In the gamma ISI model, the expectation
of the ISI is given by $\kappa\theta$. Because $\kappa$ is fixed,
$\theta$ is the input to the neuron.

Due to refractoriness, a neuron cannot fire too fast; therefore
the ISI cannot be 0 but must be larger than a few milliseconds. On the
other hand, if the ISI is too large, it means the neuron is not
working. Thus we assume the input to the neuron is trying to control
the ISI in a fixed range of time.

The average ISI, which depends on $\theta$ and $\kappa$, is limited
between $a_0$ and $b_0$, that is,
\begin{align*}
  a_{0}\le\,\overline{T}=\kappa\theta\,\le
  b_{0},~~\mbox{where}~~0<a_{0}<b_{0}<\infty.
\end{align*}
Thus, $\theta$ is bounded in
$\Omega(\kappa)=\{\theta~|~a_{0}/\kappa\le\theta\le b_{0}/\kappa\}$.

\subsection{Channel Capacity of a Single Neuron}
\label{subsec:Channel Capacity of a Single Neuron}

For a noisy channel, one important fundamental problem is to compute
the capacity $C$. Another problem is to obtain the capacity achieving
distribution.

The family of all the possible distributions $\pi(\theta)$ of inputs
$\mathcal{P}$ is defined as
\begin{align*}
  \mathcal{P}
  =
  \Bigl\{\pi~\big|~\pi(\theta)\ge0~\mathrm{for}~
  \theta\in\Omega(\kappa),
  \mbox{otherwise }0~\Bigr\}.
\end{align*}
The mutual information and the capacity depends on the choice of an
output variable. This is called ``coding'' in computational
neuroscience, but ``modulation'' is an appropriate term in information
theory. Traditionally, two types of modulations have been considered
in computational neuroscience. One is ``temporal coding'' and the
other is ``rate coding'' (see Fig.~\ref{fig:TCandRC}). Note that both
temporal and rate coding may be used in the brain. For example,
binaural sound localisation needs phase information and temporal
coding seems natural while rate coding is appropriate for a motor
neuron because muscles react according to the rate.

We provide some results on the capacity of temporal and rate coding
in the following.

\subsubsection*{Temporal Coding}

In temporal coding, the received information is $T$. For a
$\pi\in\mathcal{P}$, we define the marginal distribution as
\begin{align*}
  p(t;\pi,\kappa)=
  E_{\pi(\Theta)}[p(t|\theta;\kappa)].
\end{align*}
The mutual information between $T$ and $\Theta$ is defined as
\begin{align*}
  I(\Theta;T) &
  =E_{\pi(\Theta)}
  \biggl[
  E_{p(T|\theta;\kappa)}
  \Bigl[
    \log\frac{p(t|\theta;\kappa)}{p(t;\pi,\kappa)}
  \Bigr]
  \biggr].
\end{align*}
The capacity per spike is defined as
\begin{align*}
  C_{T} & =\sup_{\pi\in\mathcal{P}}I(\Theta;T).
\end{align*}
This optimisation problem cannot be solved analytically. However, it
has been proven that the capacity $C_T$ is achieved by a discrete
distribution with a finite number of mass points (see~\cite{Ikeda.09}
for the details).

\begin{figure}
  \centering
  \subfloat[Temporal coding ($\kappa=3$).]{
    \includegraphics[width = 0.47\figurewidth, clip]{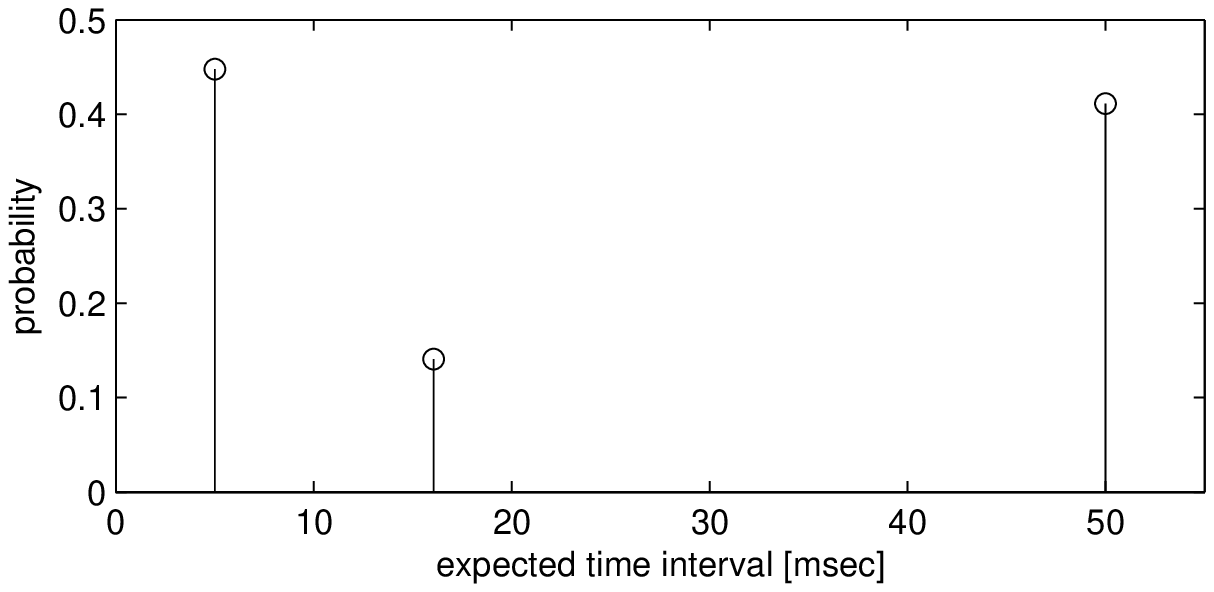}
    \label{fig:result_T}
  }
  \subfloat[Rate coding ($\kappa=3)$.]{
    \includegraphics[width = 0.47\figurewidth, clip]{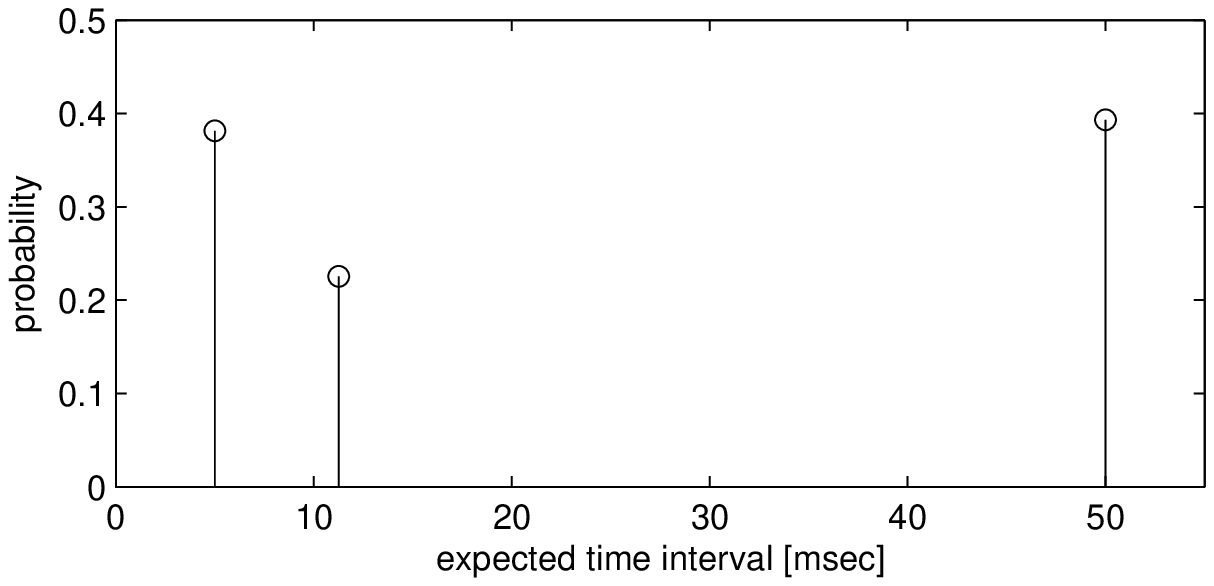}
    \label{fig:decode_R-1}
  }
  \caption{Capacity achieving distributions for temporal and rate
    coding. For both coding types, the optimal input distribution
    is discrete, with a finite number of probability mass points.}
\end{figure}

Since the optimal distribution is a discrete distribution with a
finite number of mass points, the optimisation problem becomes simple,
and we can compute the capacity and the capacity achieving
distribution numerically. Figure~\ref{fig:result_T} shows the capacity
achieving probability distribution for $\kappa=3$. The channel
capacity $C_{T}$ is 34.68bps (bit ber second) (See~\cite{Ikeda.09} for
further results).

Figure~\ref{fig:result_T} shows that the capacity is achieved when the
input is a discrete memoryless distribution\footnote{The input is
  chosen from the three states independently at each time according to
  the probability distribution shown in Fig.~\ref{fig:result_T}.} with
3 states. This does not imply that the brain is using discrete
states. It is more plausible that the brain is using continuous
states; it is likely that the actual information transmission rate in
the brain is less than the numerically computed capacity.

\subsubsection*{Rate Coding}

In rate coding, a time window is set and the number of spikes in this
interval is counted. Let us denote the interval and the rate as
$\Delta$ and $R$, respectively, and define the distribution of $R$ as
$p(r|\theta;\kappa,\Delta)$.  The form of the distribution of $R$ is
shown in~\cite{Ikeda.09}.  For $\pi\in\mathcal{P}$, let us define the
following marginal distribution
\begin{align*}
  p(r;\pi,\kappa,\Delta)
  =E_{\pi(\Theta)}[p(r|\theta;\kappa,\Delta)].
\end{align*}
The mutual information of $R$ and $\theta$ is defined as
\begin{align*}
  I(\Theta;R)
  =E_{\pi(\Theta)}
  \biggl[
  E_{p(R|\theta;\kappa,\Delta)}
  \Bigl[
    \log\frac{p(r|\theta;\kappa,\Delta)}{p(r;\pi,\kappa,\Delta)}
  \Bigr]
  \biggr].
\end{align*}
Hence, the capacity per channel use or equivalently per $\Delta$
is defined as
\begin{align*}
  C_R  =\sup_{\pi\in\mathcal{P}}I(\Theta;R).
\end{align*}
This optimisation problem cannot be solved analytically either, but
the capacity $C_{R}$ has been proven to be achievable by a discrete
distribution with a finite number of mass points~\citep{Ikeda.09}.
Figure~\ref{fig:decode_R-1} shows the capacity achieving distribution
for $\kappa=3$. The channel capacity $C_{R}$is 44.95 bits per second
(See~\cite{Ikeda.09} for further results).

\subsection{Tuning Curves}
\label{subsec:Tuning Curves}

The definition of channel capacity requires a maximisation over all
possible input probability distributions. This definition arose in an
engineering context, where a system designer is assumed to have
control over the inputs to the channel, but not the channel itself. A
different optimisation problem results if the input to the channel is
assumed to be fixed, but some control over the channel is possible.
This idea is particularly relevant for studies of biological sensory
transduction. In this context, an external stimulus that cannot be
controlled by the sensing organism must be transduced and encoded into
action potentials for communication to the brain. This stimulus can be
thought of as an input to a communication channel.

Given internal noise in the transduction mechanisms, the encoded
stimulus received by the brain is also noisy. Since we introduced
mutual information in the context of the channel coding theorem and
digital data, and here our channel input is a sensory stimulus, mutual
information may not seem relevant. However there are other reasons why
it can be useful to ensure mutual information is as large as
possible~\citep{Berger.03,Johnson.08} and we therefore are interested
in how the channel might be altered to maximise mutual information.

But what can be optimised when the stimulus cannot be controlled?  The
only other variable that can alter the mutual information is the
conditional distribution of the channel output given its input. In the
biological context this is the distribution of a neural response for a
given stimulus. Optimising this distribution means seeking to find the
communication channel that it best suited to a fixed stimulus
distribution. Without some constraints on the form of the conditional
distribution, this would not be a meaningful task. One such constraint
is to consider a fixed form for the conditional distribution that has
some parameters that can be optimised. Clearly the optimal parameter
set may change for different input distributions.

One reason for considering such an optimisation might be to assess
whether neuronal mechanisms exist for adaptively altering the
conditional distribution to match non-stationary stimuli. Another
equally intriguing reason is the idea that evolution might have
enabled neural systems to change parameters over eons with the end
result that those parameters are information theoretically optimal. In
this scenario, the underlying fixed form of the probability
distribution must be what it is due to unavoidable constraints, or
perhaps governed by criteria that are not information theoretic,
e.g. energy
considerations~\citep{Laughlin.98,Sarpeshkar,Levy.02,Laughlin.03,Berger.10}.

There are several potentially important sets of parameters that might
be chosen. However, in other contexts there has been much interest in
determining the optimal form of neuronal {\em tuning curves}, and this
is our sole focus here. Experimentally produced tuning curves are
plots of the \emph{mean} response of a neural system, as a function of
a stimulus parameter~\citep{Dayan,Lansky.08,McDonnell.PRL08}. Classic
examples of a stimulus parameter include the angle of a moving bar of
light relative to the receptive field of visual cells, or the sound
pressure level of a single frequency (pure tone) sound played by a
speaker. In these examples the average response as a function of the
stimulus defines the tuning curve. The former kind of tuning curve
typically has a bell-shape, meaning that there is a stimulus that
produces a maximal response, while more than one stimulus can produce
the same lesser response. The latter kind has a sigmoidal shape,
meaning that the mean firing rate monotonically increases with
stimulus, and here we focus only on this case.

We therefore wish to find the sigmoidal tuning curve that maximises
mutual information, for a given fixed form of a conditional
distribution. While this is generally a difficult optimisation
problem, a simple solution exists for channels where the capacity
achieving input distribution is discrete, like those considered in
this paper. In fact, the mutual information maximising tuning curve
can be derived for an arbitrary stimulus distribution, if the capacity
achieving input distribution has been calculated first. The reason for
this is explained in the following.

We consider the same noisy channel as Section~\ref{subsec:Channel
  Capacity of a Single Neuron}, that is the ISIs are governed by a
gamma distribution. We now make a slight generalisation to the setup
of Sections~\ref{subsec:Inputs and Noise of Channel} and
\ref{subsec:Channel Capacity of a Single Neuron} and consider the
expectation of the ISIs to be governed by a random variable $X$ with a
known distribution, such that $\Theta$ is an arbitrary function of
$X$.

We therefore write $\Theta=f(X)$. For the gamma ISI channel, the
tuning curve is defined as the conditional expectation of the response
variable (either $T$ or $R$) given a specified outcome of $X$. For
timing coding or rate coding respectively, we write these expectations
as
\begin{align*}
  &E_{p(T|x;\kappa)}[t]=\kappa f(x)
\end{align*}
and
\begin{align*}
  &E_{p(R|x;\kappa,\Delta)}[r]=\frac{\Delta}{\kappa f(x)}.
\end{align*}
Since $\Theta=f(X)$, when the capacity achieving distribution for
$\Theta$ is discrete with say $M$ states, the tuning curve will also
consist of $M$ unique values. Lets call these $\mu_{1}$,..,$\mu_{M}$.
While this discontinuous tuning curve achieves the largest possible
mutual information for the channel, it is not unique; other tuning
curves are equally good. Suppose $X$ is a continuous random variable,
and that the tuning curve maps large intervals of $X$ to the same $M$
values $\mu_1$, $\ldots$, $\mu_M$. This tuning curve provides the same
$M$ possibilities for the conditional expected ISI. In order for it to
provide the same mutual information achieved by the original discrete
capacity achieving distribution, it is necessary that the probability
with which each $\mu_{m}$ occurs is the same in both cases. It has
been proven for a special case of the gamma distribution and rate
coding that this can be achieved by appropriate choice for the ranges
of $X$ that are mapped to each $\mu_{m}$. The resulting optimal
discrete tuning curve is then dependent only on the probability
density function of $X$~\citep{Nikitin.PRL09}.

\begin{figure}
  \centering
  \includegraphics[width = 0.55\figurewidth]{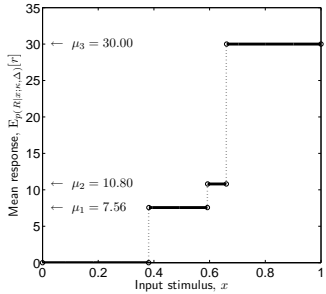}
  \caption{Channel capacity achieving tuning curve for a Poisson
    rate-coding neuron ($\kappa=1$), and an input $x\in[0,1]$, with a
    maximum spike-rate of 30 spikes per input
    sample---see~\cite{Nikitin.PRL09} for further examples.}
  \label{Fig:Tuning-1}
\end{figure}

Consequently, the capacity achieving input distributions derived in
Section~\ref{subsec:Tuning Curves} can be converted to an optimal
tuning curve for any choice of the stimulus distribution. An example
of the capacity achieving tuning curve is shown in
Figure~\ref{Fig:Tuning-1}, for the special case of $\kappa=1$, which
means the channel is equivalent to a Poisson
neuron~\citep{Nikitin.PRL09}. The maximum rate is restricted to 30
spikes per input sample.

Although such a result holds exactly only for discrete input
distributions, similar derivations of information theoretically
optimal continuous tuning curves have been made, which hold only in
the low noise
limit~\citep{McDonnell.PRL08}. See~\cite{Kostal.10,Kostal.10a} for
related work in the high noise limit.

\subsection{Discussion and Interpretation}
\label{sec:Discussion}

We have shown our results on neuron channel capacity from two very
different viewpoints. Interestingly, both show that the capacity is
achieved by a discrete distribution. The numerically computed
capacities are similar to the range indicated by some biologically
measured results of sensory neurons~\citep{Borst.99}. The channel
capacity depends on various factors, and we consider some of them
below.

\subsection{Input and Output}

We first discuss the input and output of the neuron channel in this
subsection.

Let us start with the input. Although each neuron receives information
from many neurons, we have only considered a single input $\theta$.
This may seem too simple. We assumed that the single input $\theta$
summarises all the inputs to the neuron. Moreover, $\theta$ has been
assumed to be memoryless and can have any distribution within the
support. Considering the biological system, this is far from
realistic. The net input of a neuron may not change quickly, that is,
it has memory. Moreover, a neuron's input is a collection of many
neurons' noisy outputs, therefore, it may follow a particular
probability distribution. This implies that we have computed capacity
under less restrictive assumptions, and the biologically achievable
rate should be smaller than the capacity obtained in the numerical
studies. A better understanding of the constraints on the input of a
neuron would lead to a more accurate calculation of neuronal channel
capacity.

Next, we discuss the output of a neuron from two viewpoints, decoding
and demodulation.

In order to achieve channel capacity, the receiver must act as an
optimal decoder, meaning that when $\tilde{X}$ is observed, the
receiver must compute the posterior distribution of the input $X$ as
$p(x|\tilde{x}).$ When $x$ takes discrete values $x_{1},\cdots,x_{L}$,
it becomes $p(x_{i}|\tilde{x}).$ This is a real number for each
$x_{i}$. In engineering, this type of decoding is called ``soft
decoding.''  It seems unlikely that a neuronal mechanisms for carrying
this out could exist, since a computation of the posterior distribution
is necessary.

Another standard decoding technique is ``hard decoding,'' that is,
only a single value of $x_{i}$ is chosen by the decoder. The optimal
hard decoder chooses the one which maximises the posterior
distribution, that is $\hat{x}_{i}=\mathrm{argmax}_{x_{i}}\,
p(x_{i}|\tilde{x})$. It is possible to implement the optimal hard
decoder without requiring the online computation of the posterior
distribution. Indeed, the output space (the space where $\tilde{X}$
lies) can be divided into $L$ subspaces in advance, such that the $k$
th subspace contains all the points $\tilde{X}$ such that $x_k$ has
the largest posterior probability given $\tilde{X}$. Therefore, the
optimal hard decoder can be implemented by a ``quantisation'' or
``thresholding'' algorithm which simply checks to see which subspace
$\tilde{X}$ lies in. Such an algorithm is often computationally
simpler than computing first the posterior distribution.

When we consider information processing in the brain, a naive soft
decoding seems difficult, at least for a single neuron. The hard
decoding idea seems more natural. However, the information
transmission rate of the best hard decoding is less than the best soft
decoding. We should keep this point in mind. Further discussion is
found in~\cite{Ikeda.09}.

\subsection{Discreteness}

Under the assumptions made, we have shown that the capacity of a
neuron is achieved by a discrete distribution with a finite number of
probability mass points. In information theory, there are many other
known types of channels for which channel capacity is achieved by a
discrete distribution~\citep{Huang.05}. For example, although the
capacity of an average power constrained additive Gaussian white noise
channels is achieved by continuously valued Gaussian inputs, simply
placing a constraint on the maximum amplitude of the channel means
capacity is achieved by a discrete distribution~\citep{Smith.71}.

Our results do not imply that ``neuron signals are only using discrete
levels.'' On the contrary, we believe many neurons are using
continuous levels. What is implied by our results is that those
neurons cannot achieve the capacity and the actual rate of information
transfer is therefore less than the capacity.  Another implication is
for the measurement of information capacity in
neuroscience~\citep{Borst.99}. Our result implies that only a small
number of discrete ranges are sufficient for the input distribution to
measure the information capacity of neural coding.

Information theory provides a way of quantifying whether discrete or
continuously distributed signals are better for any given
communication channel corrupted by random fluctuations. Many
neuroscience studies quantify information using Shannon's famous
information \textsl{capacity} formula relating mutual information to
signal-to-noise ratio. This formula is correct only for Gaussian
additive noise channels~\citep{Cover2} --- it relies on many
assumptions, and if any are false it can significantly under- or
over-estimate the true capacity~\citep{Berger.98}.

\subsection{Further Points}

Guessing the channel model is not straightforward. In particular,
feedback is prevalent in many parts of the brain, which makes it much
more difficult to relate changes in responses to inputs%
\footnote{Channels with feedback can have larger capacity.}. One
important example where it is known that the circuitry is solely
feedforward is that of retinal cells~\citep{Jacobs.09}. This example
has been used to demonstrate that it is possible to rule out certain
guesses of neural codes. Because this is based on optimal Bayesian
decoding of a forced binary choice, it would be interesting to
extend~\citep{Jacobs.09} beyond the binary limitation to that where
the signals being coded may have many possibilities

What can we say about more complicated situations? For example,
information processing of cortical neurons are not strictly
feedforward and the information is shared by many neurons. If we
assume every neuron is performing the same computation, and each
neuron encodes and decodes information in the same way, it seems
possible to extend our results. However, when different neurons encode
information differently, the problem becomes very
different. Understanding how the brain works in information theoretic
terms is one of the grand challenges of this century.

\section{Conclusions and Future Directions}\label{S:conclusion}

Most information theory research to date has been predicated on an
engineering viewpoint. The main thrust has been to \emph{design}
compression/decompression methods that compress source information to
the limits given by its entropy (source coding), or to \emph{design}
error correction schemes and encoding/decoding methods that allow
communication close to capacity.

On the other hand, the goal of neuroscience is instead to
\textit{understand} information processing in the brain.

This does not mean that information theory has no place in
computational neuroscience. Information theory provides a way
to\textit{ measure} information and to understand the \emph{limits} of
compression (namely, entropy) and communication (namely, mutual
information and channel capacity).

A common ``information theoretic'' method in computational
neuroscience is to obtain quantitative estimates of the mutual
information between observed sets of data. Since various methods for
the (difficult) problem of accurately estimating mutual information
exist, the bigger difficulty is with using the result to say something
about how the system works. Indeed, the actual goal of computational
neuroscience is that of ``system identification,'' as engineers would
call it.

If the brain uses spikes to transmit information (which on faster
time-scales appears to be the case) then understanding the neural code
--- how the brain encodes information before sending it across the
``channel'' --- is tantamount to understanding how the brain
works. Indeed, if we would listen in to the messages as they are sent
from one neuron to another, it would be relatively straightforward to
determine what each neuron is doing.

Although system identification is not the forte of traditional
information theory~\citep{Johnson.08ITW}, this is merely for
historical reasons. With computational neuroscience as a main
motivator, we predict that the next decade will see the expansion of
information theory to include more powerful techniques for system
identification, and possibly even an integration of control,
computation and information theory into a unified framework.

Some recent information theoretic approaches in neuroscience that go
beyond standard Shannon theory are summarized in the following list.

\begin{itemize}
\item While it is traditional in engineering to separately code an
  information source, and then {\em channel code} it for communication
  across a channel, it has been shown for some simple but instructive
  examples, that non-separation of these two aspects can achieve an
  optimal communication system with a vastly reduced complexity
  compared to separation~\citep{Gastpar.03}. This fact is potentially
  important for neurobiological systems, where separation mechanisms
  seem implausible.
\item Many studies have investigated whether the brain might have
  mechanisms for implementing Bayesian algorithms during decision
  making, prediction and pattern
  recognition~\citep{Rao.99,Lee.03,Knill.04,George.09}.
\item The possibility of analog cortical error correcting codes has
  been proposed by~\cite{Fiete.08}.
\item One limitation of Shannon theory is that its measures say
  nothing about directionality or causality. However, directed
  information theory has also been
  developed~\citep{Granger.69,Marko.73,Rissanen.87,Massey.90,Tatikonda.09}
  and has recently been applied in
  neuroscience~\citep{Hesse.03,Eichler.06,Waddell.07,Amblard.11,Quinn.11}.
\item The relationship between control, information theory and
  thermodynamics has been discussed by~\cite{Mitter.05,Friston.10,Mitter.10}.
\end{itemize}

Summarising our own modest contribution, we carefully came up with a
simple neuron channel model based on biological evidence and computed
the channel capacity of this model. Interestingly, it was proved that
the channel capacity is achieved by a discrete distribution.

\section*{Acknowledgements}
  Mark D. McDonnell's contribution was supported by the Australian
  Research Council (ARC) under ARC grants DP1093425~(including an
  Australian Research Fellowship), RN0459498 and DP0770747. Jonathan
  H. Manton wishes to acknowledge several interesting and
  thought-provoking conversations with Professor Iven Mareels which
  facilitated the writing of parts of this paper. JM's contribution
  was supported in part by the Australian Research Council.


\end{document}